# CESR Test Accelerator

August 2013

David L. Rubin,
Cornell University

The Cornell Electron Storage Ring (CESR) was reconfigured as the CESR Test Accelerator (CesrTA) in 2008[1] following the conclusion of the three decade long electron-positron colliding beam program. During the past five years, CesrTA has made and continues to make important contributions to our understanding of beam physics in the ultra-low emittance and high charge density regime characteristic of damping rings for electron positron colliders. The CESR ring is instrumented with; (a)detectors developed especially for measuring the electron cloud that is evolved from the walls of the vacuum chambers by synchrotron radiation, (b)precision beam size monitors for measuring both horizontal and vertical beam size of every single bunch on every turn in a long train of bunches, and (c)one hundred high bandwidth beam position monitors that provide turn-by-turn measurement of each bunch in a train. In addition, CESR is equipped with a dozen superconducting damping wigglers that increase the radiation damping rate by an order of magnitude[2]. CESR is truly a wiggler dominated damping ring.

Approximately 40 days/year of storage ring running time are dedicated to the beam physics experimental program. (The accelerators are operated for the synchrotron light source CHESS during the remainder of the year.) The flexibility of the storage ring optics, and the integration through the control system with sophisticated accelerator modeling software, has made possible an extraordinarily broad range of experiments. Graduate and undergraduate students, postdocs, and accelerator staff scientists have made measurements, analyzed data, designed and built instrumentation, and documented their work in conference proceedings, journal articles, and PhD thesis[3,4,5]. Program graduates have expertise in lattice design and optimization, the physics of low emittance electron/positron beams, and perhaps most importantly, machine operation and beam based measurements.

Some aspects of the research are described in more detail below.

### Electron Cloud Growth

The 30 retarding field analyzers(RFAs) in CESR measure time averaged position and energy distribution of the electrons generated in the magnetic fields of the four principle guide field elements, namely dipole, quadrupole, wiggler, and field free (drift). RFAs installed in chambers with different mitigations provide direct comparison of their efficacy[6]. All of the mitigations proposed for the ILC damping rings have been tested in CesrTA[7]. Planning is underway to test chambers being considered for the CLIC damping ring, where antechambers are designed to capture > 99% of the synchrotron radiation photons.

Shielded pickups (SPU) allow a direct measurement of the temporal development and decay of the cloud, providing a powerful test of our electron cloud physics models[8]. We have also developed detectors with the combined capability of both RFA (spatial and energy) and SPU(temporal)[9]. The studies of the growth and mitigation of the electron cloud, have been critical to the design of the vacuum system for Super KEK B facility as well as the ILC damping rings. Electron cloud mitigations under consideration for the LHC upgrade have been tested at CesrTA. Physics models of the growth and decay of the cloud have been benchmarked with measurements at CesrTA.

### Electron Cloud Emittance Dilution and Instabilities
The storage ring is instrumented to measure the emittance dilution as well as instabilities induced by the electron cloud. Bunch by bunch measurements of both vertical and horizontal emittance provide a direct measurement of bunch dependent emittance growth. From bunch by bunch position spectra we extract information about bunch dependent instabilities[10].

### Emittance tuning
A sequence of beam based measurements developed at CesrTA, for identifying sources of emittance diluting optical errors and then compensating those sources, routinely yields vertical geometric emittance $\varepsilon_v$< 10pm-rad[3,11]. The procedure is executed in just a few minutes. The ability to quickly establish ultra-low vertical emittance in a variety of optical configurations has been crucial to the study of emittance diluting collective effects such as the electron cloud, intra-beam scattering, as well as single particle effects.

As we continue to identify and eliminate sources of single particle vertical emittance, our sensitivity to more subtle effects necessarily increases. The residual emittance is presently dominated by non-static sources such as noise coupled to the beam through magnet power supplies, RF cavities, and/or feedback kickers. Our goal is to reduce these external contributions to the beam size, and to achieve the quantum limited vertical emittance (~0.2 pm-rad).

### Intra-beam scattering
The capability to simultaneously monitor horizontal, vertical, and longitudinal emittances of a single bunch permits direct measurement of the dependence of bunch size on bunch charge. We have exploited this capability to measure the dependence of the intra-beam scattering growth rate on beam energy, zero current emittance, RF voltage, RF cavity dispersion, radiation damping time, and transverse coupling[12]. The CesrTA measurements and analysis of intra-beam scattering in electron/positron beams are the most complete to date. Meanwhile, our sensitivity will only increase as we continue to eliminate the sources of the residual (single particle) vertical emittance.

### Fast Ion Instability

The fast ion instability that appears in a train of electron bunches, is the approximate analog of the electron cloud instability in a train of positrons. Ionization of the residual gas results in a cloud of positively charged ions that are trapped in the potential well of the electron beam. The ions couple motion of leading to trailing bunches that can lead to instability. CESR is uniquely instrumented to identify signatures of the instability through the bunch by bunch positron spectra and to measure the bunch dependent emittance growth. Measurements of the fast ion instability are planned for an upcoming machine studies period.

### ILC Damping Ring

Findings at CesrTA with respect to electron cloud effects, emittance tuning techniques, and beam instrumentation for measuring electron cloud, beam sizes, and beam positions are the basis for much of the design of the ILC damping rings as documented in the ILC-Technical Design Report[13]. The studies of intra-beam scattering and fast ion instabilities are more relevant to the design of the CLIC damping rings, which have somewhat more ambitious specifications. The program has allowed the Cornell group to cultivate the kind of talent that will be absolutely essential to the final engineering design, and commissioning of the damping rings for a linear collider. The Cornell group has ambitions to make a major contribution to the realization of the damping rings for the ILC.